\documentclass{appolb}
\usepackage{epsfig}

\begin{document}
\title{Moment equations in a Lotka-Volterra extended system
with time correlated noise%
\thanks{Presented at the $19^{th}$ Marian Smoluchowski Symposium on Statistical Physics,
Krak\'ow, Poland, May 14-17, 2006
}%
}
\author{D. Valenti$^a$\footnote{e-mail: valentid@gip.dft.unipa.it},
L. Schimansky-Geier$^b$\footnote{e-mail: alsg@physik.hu-berlin.de},
X. Sailer$^b$, B. Spagnolo$^a$ and M. Iacomi$^{a,c}$
\address{$^a$Dipartimento di Fisica e Tecnologie Relative, Universit\`a di
Palermo\\
and CNISM - Unit\`a di Palermo, Group of Interdisciplinary Physics\\
Viale delle Scienze, I-90128 Palermo, Italy\\
http://gip.dft.unipa.it \medskip
\\$^b$Institut f\"{u}r
Physik, Humboldt Universit\"{a}t zu
Berlin\\
Newtonstra{\ss}e 15, D-12489 Berlin, Germany\\
http://summa.physik.hu-berlin.de\\ \medskip $^c$Institutul de
\c{S}tiin\c{t}e Spa\c{t}iale \\ P.O.Box MG-23, Ro 76900,
Bucharest-M\u{a}gurele, Romania\\
http://www.spacescience.ro}}
 \maketitle
\begin{abstract}
A spatially extended Lotka-Volterra system of two competing species
in the presence of two correlated noise sources is analyzed: (i) an
external multiplicative time correlated noise, which mimics the
interaction between the system and the environment; (ii) a
dichotomous stochastic process, whose jump rate is a periodic
function, which represents the interaction parameter between the
species. The moment equations for the species densities are derived
in Gaussian approximation, using a mean field approach. Within this
formalism we study the effect of the external time correlated noise
on the ecosystem dynamics. We find that the time behavior of the
$1^{st}$ order moments are independent on the multiplicative noise
source. However the behavior of the $2^{nd}$ order moments is
strongly affected both by the intensity and the correlation time of
the multiplicative noise. Finally we compare our results with those
obtained studying the system dynamics by a coupled map lattice
model.
\end{abstract}




\section{Introduction}
Real ecosystems are influenced by the presence of continuous
external fluctuations connected to the random variation of
environmental parameters such as temperature and natural resources,
which affects the system dynamics by a multiplicative nonlinear
interaction~\cite{Zimmer}. The spatio-temporal behaviour and the
formation of spatial patterns became recently an important topic in
hydrodynamics systems, nonlinear optics, oscillatory chemical
reactions and in theoretical ecology~\cite{Cross}-\cite{Sinclair}.
During the last three decades a new approach, based on the use of
moments, has been exploited to describe the behaviour of spatially
extended system in population dynamics, quantum systems in the
context of nonlinear Schr\"odinger equations, and kinetic models of
polymer dynamics~\cite{Bolker}-\cite{Ilg}. In this paper, by using
the formalism of the moments, we study the spatio-temporal behaviour
of a two-dimensional system formed by two competing species subject
to random fluctuations. The system is described by generalized
Lotka-Volterra equations in the presence of two noise sources: (i) a
multiplicative time correlated noise with correlation times
$\tau^c$, modeled as an Ornstein-Uhlenbeck process~\cite{Gardiner},
which takes the environment fluctuations acting on the species into
account; (ii) a noisy interaction parameter which is a stochastic
process, whose dynamics is given by a periodic function in the
presence of a correlated dichotomous noise, with correlation time
$\tau_d$. We define a two-dimensional spatial domain considering in
each site a system of two Lotka-Volterra equations coupled by an
interaction term~\cite{Lotka}. Afterwards, using a mean field
approach, we study the dynamics of the system by the moment
equations, within the Gaussian approximation~\cite{Kawai,Valenti1},
getting the time behavior of the $1^{st}$ and $2^{nd}$ order moments
of the species concentrations. Finally we compare our results with
those obtained within the formalism of the coupled map lattice (CML)
model~\cite{Kaneko}.

\section{The model}
\vskip-0.2cm Our system is described by a time evolution model of
Lotka-Volterra equations, within the Ito
scheme~\cite{Valenti2}-\cite{Valenti3}, with diffusive terms in a
spatial lattice with $N$ sites
\begin{eqnarray}
\dot{x}_{i,j}&=&\mu x_{i,j} (1-x_{i,j}-\beta y_{i,j})+x_{i,j}
\sqrt{\sigma_x}\zeta^x_{i,j} + D\sum_\gamma (x_\gamma-x_{i,j})
\label{Lotka_eq_1}\\
\dot{y}_{i,j}&=&\mu y_{i,j} (1-y_{i,j}-\beta x_{i,j})+ y_{i,j}
\sqrt{\sigma_y}\zeta^y_{i,j} + D\sum_\gamma (y_{\gamma}-y_{i,j}),
\label{Lotka_eq_2}
\end{eqnarray}
where $x_{i,j}$ and $y_{i,j}$ denote respectively the densities of
species $x$ and species $y$ in the lattice site $(i,j)$, $\mu$ is
the growth rate, $D$ is the diffusion constant, and $\sum_\gamma$
indicates the sum over all the sites except the pair $(i,j)$. Here
$\beta$ is the interaction parameter. $\zeta^l(t)$ $(l=x,y)$ are
statistically independent colored noises, i. e. exponentially
correlated processes given by the Ornstein-Uhlenbeck
process~\cite{Gardiner}
\begin{equation}
\frac{d\zeta^l}{dt}=-\frac{1}{\tau^c_l}\zeta^l + \frac{1}{\tau^c_l}
\xi^l(t) \qquad (l=x,y) \label{colored_noise}
\end{equation}
and $\xi^l(t)$ $(l=x,y)$ are Gaussian white noises within the Ito
scheme with zero mean and correlation function $\langle
\xi^l(t)\xi^m(t')\rangle = 2\sigma \delta(t-t')\delta_{lm}$. The
correlation function of the processes of Eq.($\ref{colored_noise}$)
is
\begin{equation}
\langle \zeta^l(t)\zeta^m(t')\rangle = \frac{\sigma}{\tau^c_l}
e^{-|t-t'|/\tau^c_l} \delta_{lm}
 \label{correlation function}
\end{equation}
and gives $2\delta(t-t')\delta_{lm}$ in the limit $\tau^c_l
\rightarrow 0$.
\subsection{The interaction parameter}
\vskip-0.2cm

The value of the interaction parameter $\beta$ is crucial for the
dynamical regime of the ecosystem investigated. In fact, for $\beta
< 1$ both species survive and a coexistence regime takes place,
while for $\beta > 1$ one of the two species extinguishes after a
certain time and exclusion occurs. These two regimes correspond to
stable states of the Lotka-Volterra's deterministic
model~\cite{Valenti2}. Moreover periodical and random driving forces
connected with environmental and climatic variables, such as the
temperature, modify the dynamics of the ecosystem, affecting both
directly the species densities and the interaction parameter. This
causes the system dynamics to change between coexistence ($\beta <
1$) and exclusion ($\beta > 1$) regimes. To describe this dynamical
behavior we consider as interaction parameter $\beta(t)$ a
dichotomous stochastic process, whose jump rate is a periodic
function
%
\begin{eqnarray}
\gamma(t) &=& 0, \qquad\qquad\qquad\qquad\enspace  \Delta t \leq \tau_d\nonumber\\
\label{jump_rate}\\
\gamma(t) &=& \gamma_0 \left(1 + A \thinspace \vert \cos\omega t
\vert \right), \quad \Delta t > \tau_d\;.\nonumber
\end{eqnarray}

\begin{figure}[htbp]
\begin{center}
\includegraphics[width=5.5cm]{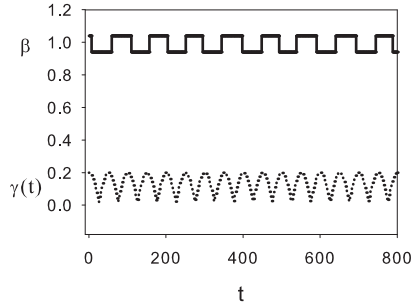}
\end{center}
\vskip-0.6cm \caption{\small Time evolution of the interaction
parameter $\beta(t)$, driven by time correlated dichotomous noise
with correlation time $\tau_d = 43.5$. $\beta(t)$ jumps between
$\beta_{down}=0.94$ and $\beta_{up}=1.04$. The initial value is
$\beta(0)=\beta_{up}=1.04$. The values of the other parameters are:
$A =9.0$, $\omega/(2\pi)=10^{-2}$,
$\gamma_0=2\cdot10^{-2}$.\bigskip}
 \label{beta}
\vskip-0.5cm
\end{figure}

%
Here $\Delta t$ is the time interval between two consecutive
switches, and $\tau_d$ is the delay between two jumps, that is the
time interval after a switch, before another jump can occur. In
Eq.~(\ref{jump_rate}), $A$ and $\omega = (2\pi)/T$ are respectively
the amplitude and the angular frequency of the periodic term, and
$\gamma_0$ is the jump rate in the absence of periodic term. Setting
$\beta_{down}=0.94<1$ and $\beta_{up}=1.04>1$, the dichotomous noise
causes $\beta(t)$ to jump between two values, $\beta_{down}$ and
$\beta_{up}$. The value $\tau_d = 43.5$ for the delay corresponds to
a competition regime with $\beta$ switching quasi-periodically from
coexistence to exclusion regimes~\cite{Valenti1} (see
Fig.~\ref{beta}).
This synchronization phenomenon is due to the choice of the $\tau_d$
value, which stabilizes the jumps in such a way they happen for high
values of the jump rate, that is for values around the maximum of
the function $\gamma(t)$. This causes a quasi-periodical time
behaviour of the species concentrations $x$ and $y$, which can be
considered as a signature of the stochastic resonance
phenomenon~\cite{Benzi}.
\section{Mean field model}
\vskip-0.2cm

\begin{figure}[htbp]
\begin{center}
\includegraphics[width=8.5cm]{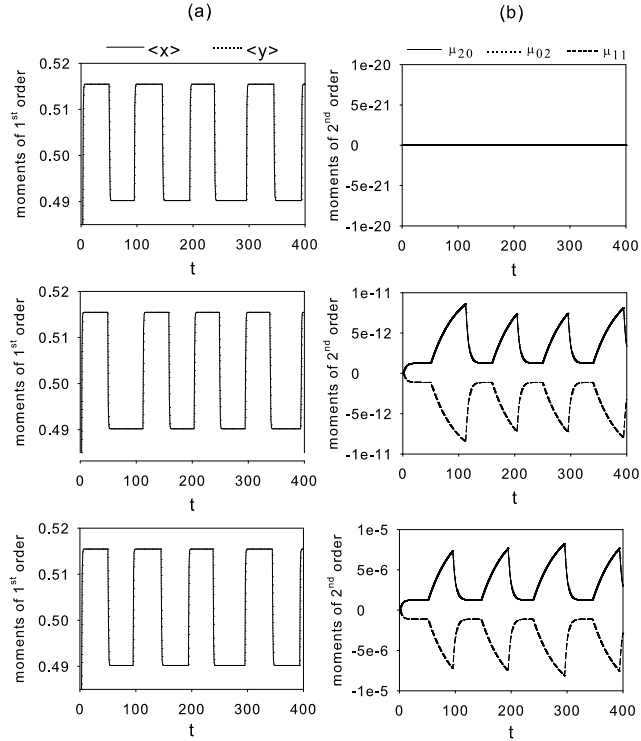}
\end{center}
\vskip-0.3cm\caption{\small Time evolution of the $1^{st}$ and
$2^{nd}$ order moments for $\tau^c=1$. The time series of (a)
$<x(t)>$ and $<y(t)>$, and (b) $\mu_{2,0}$ and $\mu_{0,2}$
respectively, are completely overlapped. The values of the
multiplicative noise intensity are: $\sigma=0, 10^{-12}, 10^{-6}$,
from top to bottom. Here $\tau_d = 43.5$, $\mu=2$, and $D=0.05$. The
initial conditions are: $\zeta^x (0) = \zeta^y (0) = 0$, $<x(0)>$ =
$<y(0)>$ = $0.1$, $\mu_{20}(0)=\mu_{02}(0)=\mu_{11}(0)=0$. The
values of the other parameters are the same of Fig.~\ref{beta}.}
 \label{moment_series1}
 \vskip-0.3cm
\end{figure}

\begin{figure}[htbp]
\begin{center}
\includegraphics[width=8.5cm]{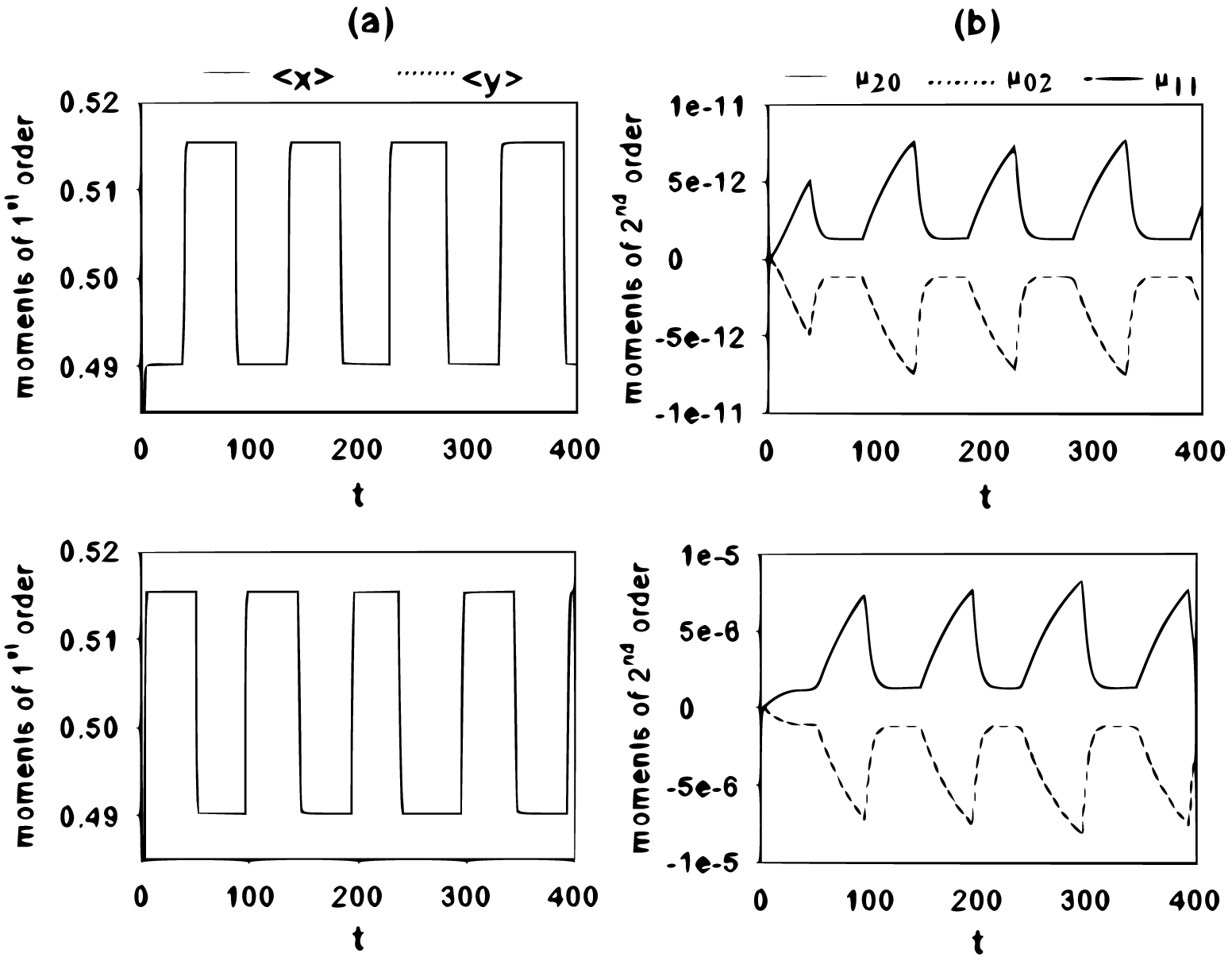}
\end{center}
\vskip-0.3cm\caption{\small Time evolution of the $1^{st}$ and
$2^{nd}$ order moments for $\tau^c=10$. Here (a) $<x(t)>$ and
$<y(t)>$, and (b) $\mu_{2,0}$ and $\mu_{0,2}$ respectively, are
overlapped. The values of the multiplicative noise intensity are:
$\sigma=10^{-12}, 10^{-6}$, from top to bottom. The values of the
other parameters and the initial conditions are the same of
Fig.~\ref{moment_series1}.}
 \label{moment_series2}
\end{figure}

\begin{figure}[htbp]
\begin{center}
\includegraphics[width=8.5cm]{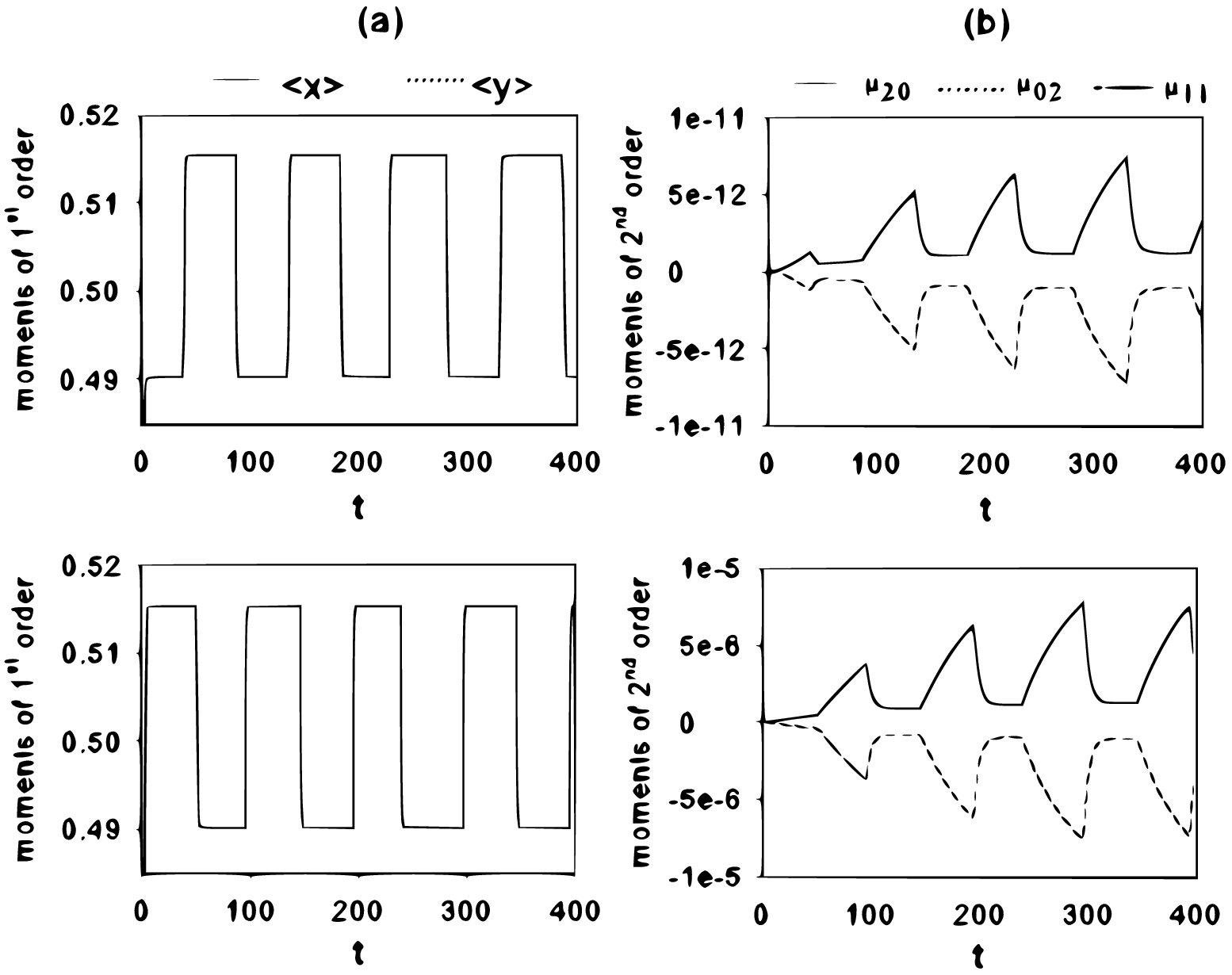}
\end{center}
\vskip-0.3cm\caption{\small Time evolution of the $1^{st}$ and
$2^{nd}$ order moments for $\tau^c=100$. Here (a) $<x(t)>$ and
$<y(t)>$, and (b) $\mu_{2,0}$ and $\mu_{0,2}$ respectively, are
overlapped. The values of the multiplicative noise intensity are:
$\sigma=10^{-12}, 10^{-6}$, from top to bottom. The values of the
other parameters and the initial conditions are the same of
Fig.~\ref{moment_series1}.}
 \label{moment_series3}
\end{figure}

In this section we derive the moment equations for our system.
Assuming $N \rightarrow \infty$, we write Eqs.~(\ref{Lotka_eq_1})
and (\ref{Lotka_eq_2}) in mean field form
\begin{eqnarray}
\dot{x}&=&f_x(x,y)+\sqrt{\sigma_x} g_x(x) \zeta^x + D(<x>-x)
\label{mean_eq_1}\\
\dot{y}&=&f_y(x,y)+\sqrt{\sigma_y} g_y(y) \zeta^y + D(<y>-y),
\label{mean_eq_2}
\end{eqnarray}
where $<x>$ and $<y>$ are average values on the spatial lattice
considered, that is the ensemble average in the thermodynamics
limit. We set $f_x(x,y)=\mu x (1-x-\beta y)$, $g_x(x)=x$,
$f_y(x,y)=\mu y (1-y-\beta x)$, $g_y(y)=y$. By site averaging
Eqs.~(\ref{mean_eq_1}) and (\ref{mean_eq_2}), we obtain
\begin{eqnarray}
<\dot{x}>~=~<f_x(x,y)>,
\label{mean site eq1}\\
<\dot{y}>~=~<f_y(x,y)>. \label{mean site eq2}
\end{eqnarray}
By expanding the functions $f_x(x,y)$, $g_x(x)$, $f_y(x,y)$,
$g_y(y)$ around the $1^{st}$ order moments $<x>$ and $<y>$, we get
an infinite set of simultaneous ordinary differential equations for
all the moments. This equation set is truncated by using a Gaussian
approximation, which causes the cumulants above the $2^{nd}$ order
to vanish. Therefore we obtain
\begin{eqnarray}
<\dot{x}>&=&\mu <x> (1-<x>-\beta <y>) - \mu(\beta\mu_{11}+\mu_{20})
\label{x_mean_eq}\\
<\dot{y}>&=&\mu <y> (1-<y>-\beta <x>) -
\mu(\beta\mu_{11}+\mu_{02}) \label{y_mean_eq}\\
\dot{\mu}_{20}&=&2\mu \mu_{20} - 2D\mu_{20} - 2\mu\beta<y>\mu_{20}
- 2\mu <x>(\beta\mu_{11}+2\mu_{20})\nonumber\\
&+& 2\sigma_x(<x>^2+\mu_{20})(1-e^{-\frac{t}{\tau_x^c}})\label{mu20_eq}\\
\dot{\mu}_{02}&=&2\mu \mu_{02} - 2D\mu_{02} - 2\mu\beta<x>\mu_{02}
- 2\mu <y>(\beta\mu_{11}+2\mu_{02})\nonumber\\
&+& 2\sigma_y(<y>^2+\mu_{02})(1-e^{-\frac{t}{\tau_y^c}})\label{mu02_eq}\\
\dot{\mu}_{11}&=&2\mu\mu_{11} - 2D\mu_{11}
-<x>[2\mu\mu_{11}+\mu\beta(\mu_{11}+\mu_{02})]\nonumber\\
&-&<y>[2\mu\mu_{11}+\mu\beta(\mu_{11}+\mu_{20})]~,
 \label{mu11_eq}
\end{eqnarray}
where $\mu_{20}$, $\mu_{02}$, $\mu_{11}$ are the $2^{nd}$ order
central moments defined on the lattice
\begin{eqnarray}
\mu_{20}&=&<x^2>-<x>^2, \qquad \mu_{02} = <y^2>-<y>^2 \label{mu20}, \\
\mu_{11}&=&<xy>-<x><y>~.\label{mu11}
\end{eqnarray}
The dynamics of the two species is analyzed through the time
behavior of $1^{st}$ and $2^{nd}$ order moments according to
Eqs.~(\ref{x_mean_eq})-(\ref{mu11_eq}), for different values both of
intensities $\sigma$ ($=\sigma_x=\sigma_y$) and correlation time
$\tau^c$ (=$\tau_x^c = \tau_y^c$) of the multiplicative colored
noise. The results are reported in
Figs.~\ref{moment_series1}-\ref{moment_series3}. The results shown
in Fig.~\ref{moment_series1}, obtained with a low value of the
correlation time $\tau^c$ reproduce the same time behavior obtained
for uncorrelated white noise sources in Ref.~\cite{Valenti1}. The
value of the time delay is $\tau_d = 43.5$, which determines a
quasi-periodic switching between the coexistence and exclusion
regimes. The values of the other parameters are: $\mu=2$, $D=0.05$.
The initial conditions are: $\zeta^x (0) = \zeta^y (0) = 0$,
$<x(0)>$ = $<y(0)>$ = $0.1$, $\mu_{20}(0)= \mu_{02}(0) =
\mu_{11}(0)= 0$. These initial values for the moments correspond to
uniformly distributed species on the considered lattice. In
Figs.~\ref{moment_series1}a,~\ref{moment_series2}a,~\ref{moment_series3}a
we note that the $1^{st}$ order moments undergo correlated
oscillations around $0.5$. This behaviour is independent both on the
intensity and the correlation time of the multiplicative noise. On
the other hand the behavior of the $2^{nd}$ order moments depends
strongly both on the intensity and the correlation time of the
external multiplicative noise. In the absence of noise $\mu_{20}$,
$\mu_{02}$, $\mu_{11}$ maintain their initial values. For very low
levels of multiplicative noise ($\sigma=10^{-12}$) quasi-periodical
oscillations appear with the same frequency of the interaction
parameter $\beta (t)$, because the noise breaks the symmetry of the
dynamical behavior of the $2^{nd}$ order moments (see
Figs.~\ref{moment_series1}b,~\ref{moment_series2}b,~\ref{moment_series3}b).
About the variances of the two species, the time series of
$\mu_{20}$ and $\mu_{02}$, which coincide all the time, show an
oscillating behaviour characterized by small (close to zero) and
large values. However the negative values of the correlation
$\mu_{11}$ indicate that the two species distributions are
anticorrelated. In particular, we find a time behavior characterized
by anticorrelated oscillations, whose amplitude increases with the
multiplicative noise intensity and it is reduced as the correlation
time becomes bigger (see
Figs.~\ref{moment_series1}b,~\ref{moment_series2}b,~\ref{moment_series3}b).
This anticorrelated behaviour indicates that the spatial
distribution in the lattice will be characterized by zones with a
maximum of concentration of species $x$ and a minimum of
concentration of species $y$ and viceversa. The two species will be
distributed therefore in non-overlapping spatial patterns. This
physical picture is in agreement with previous results obtained with
a different model~\cite{Valenti3}. A further increase of the
multiplicative noise intensity ($\sigma = 10^{-6}$) causes an
enhancement of the oscillation amplitude both in $\mu_{20}$,
$\mu_{02}$ and $\mu_{11}$. This gives information on the probability
density of both species, whose width and mean value undergo the same
oscillating behavior. The anticorrelated behavior is enhanced by
increasing the noise intensity value. We note that the amplitude of
the oscillations is of the same order of magnitude of the noise
intensity $\sigma$, that is the amplitude of the oscillations is
enhanced as the noise intensity increases. The periodicity of these
noise-induced oscillations shown in
Figs.~\ref{moment_series1}-\ref{moment_series3} is the same of the
interaction parameter $\beta(t)$ (see Fig.~\ref{beta}). Moreover the
right-hand side of Figs.~\ref{moment_series1}-\ref{moment_series3}
shows that the multiplicative colored noise affects the time
evolution of the $2^{nd}$ order moments introducing a delay: the
amplitude of the oscillations reaches its highest value after a time
interval whose length increases as $\tau^c$ becomes bigger. In fact
by comparing the first figures in
Figs.~\ref{moment_series1}b-\ref{moment_series3}b with $\sigma =
10^{-12}$, for example, the maximum is reached approximately at
$t\simeq110$ for $\tau^c = 1$, and at $t\simeq330$ for $\tau^c =
100$.

\section{Coupled Map Lattice Model}
\vskip-0.2cm

Our results, obtained by using moment equations in Gaussian
approximation, can be checked studying the dynamics of the system by
a different approach, namely the CML model
\begin{eqnarray}
x_{i,j}^{(n+1)}&=&\mu x_{i,j}^{(n)} (1-x_{i,j}^{(n)}-\beta^{(n)}
y_{i,j}^{(n)})+\sqrt{\sigma_x} x_{i,j}^{(n)} \zeta_{i,j}^{x{(n)}} +
D\sum_\gamma (x_{\gamma}^{(n)}-x_{i,j}^{(n)})\enspace\qquad
\label{CLM-Lotka_1}\\
y_{i,j}^{(n+1)}&=&\mu y_{i,j}^{(n)} (1-y_{i,j}^{(n)}-\beta^{(n)}
x_{i,j}^{(n)})+\sqrt{\sigma_y} y_{i,j}^{(n)} \zeta_{i,j}^{y{(n)}} +
D\sum_\gamma (y_{\gamma}^{(n)}-y_{i,j}^{(n)}),\enspace\qquad
\label{CLM-Lotka_2}
\end{eqnarray}
which represents a discrete version of the Lotka-Volterra equations.
\begin{figure}[htbp]
\begin{center}
\includegraphics[width=8.5cm]{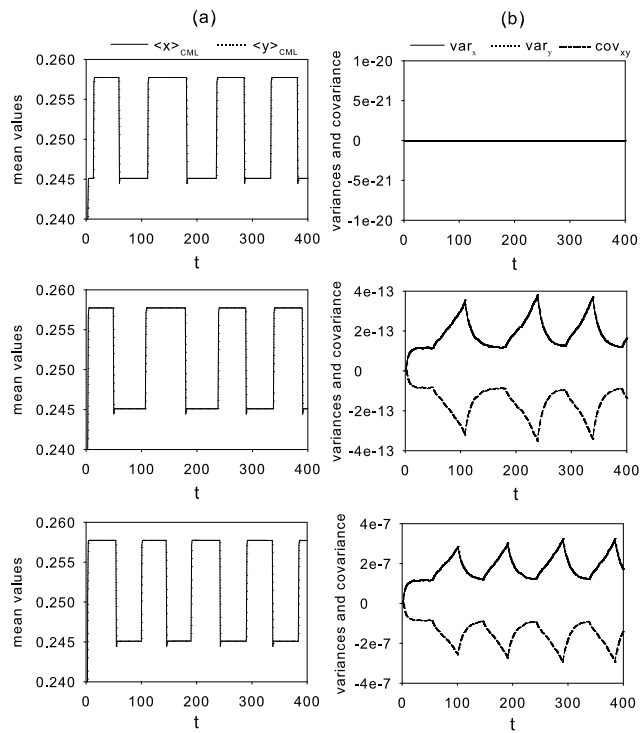}
\end{center}
\vskip-0.3cm\caption{ \small (a) Mean values: $<x>_{_{CML}}$,
$<y>_{_{CML}}$, and (b) variances: $var_x$, $var_y$ and $cov_{xy}$
of the two species, as a function of time for $\tau^c=1$. The values
of the multiplicative noise intensity are: $\sigma=0, 10^{-12},
10^{-6}$, from top to bottom. The initial values of the species
concentrations are $x_{i,j}^{(0)} = y_{i,j}^{(0)} = 0.1$ for all
sites $(i,j)$. The values of the other parameters are the same of
Fig.~\ref{moment_series1}.}
 \label{CML_series1}
\end{figure}

\begin{figure}[htbp]
\begin{center}
\includegraphics[width=8.5cm]{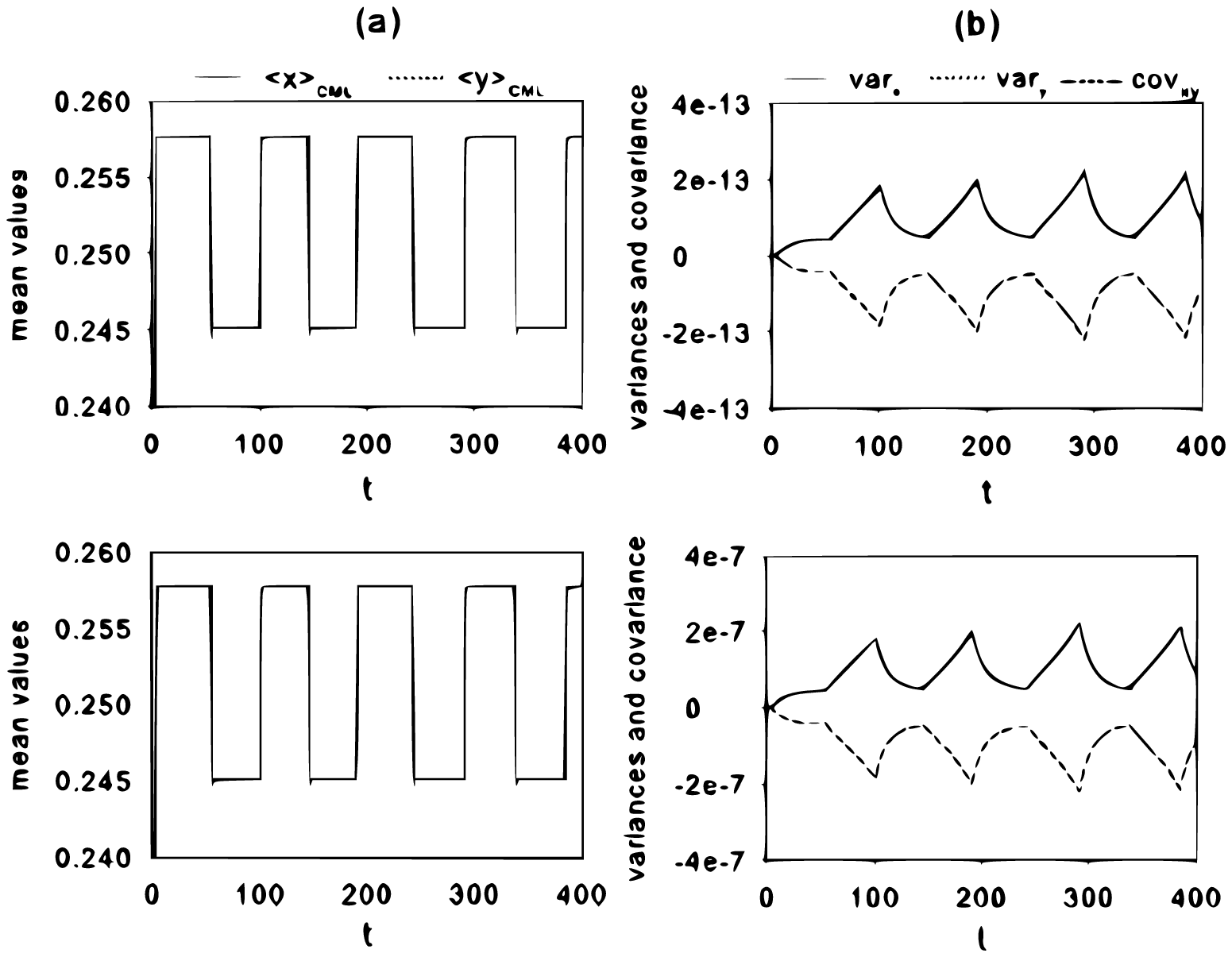}
\end{center}
\vskip-0.3cm\caption{\small (a) Mean values: $<x>_{_{CML}}$,
$<y>_{_{CML}}$, and (b) variances: $var_x$, $var_y$ and $cov_{xy}$
of the two species, as a function of time for $\tau^c=10$. The
values of the multiplicative noise intensity are: $\sigma=10^{-12},
10^{-6}$, from top to bottom. The values of the other parameters and
the initial conditions are the same of Fig.~\ref{CML_series1}.}
 \label{CML_series2}
\end{figure}

\begin{figure}[htbp]
\begin{center}
\includegraphics[width=8.5cm]{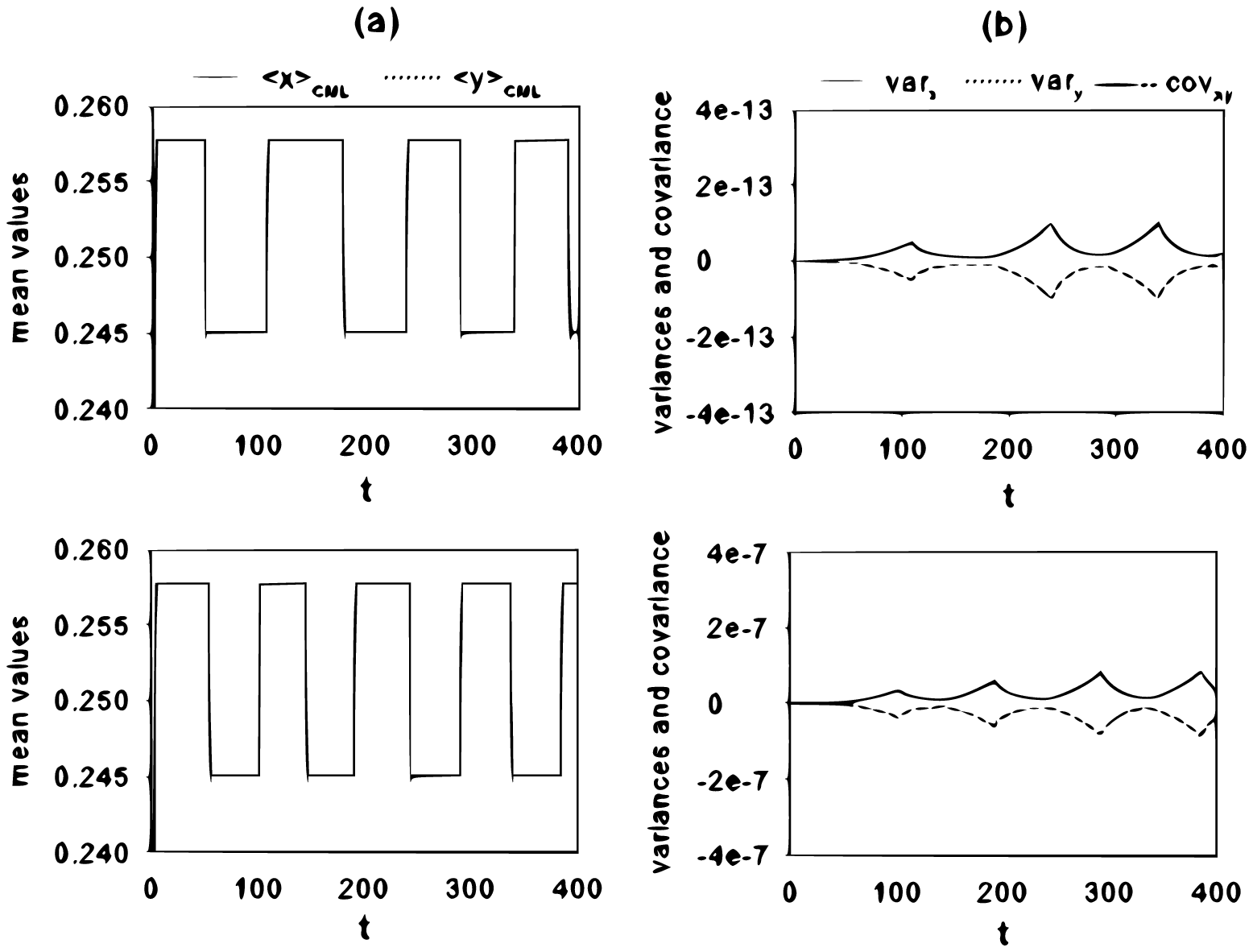}
\end{center}
\vskip-0.3cm\caption{\small (a) Mean values: $<x>_{_{CML}}$,
$<y>_{_{CML}}$, and (b) variances: $var_x$, $var_y$ and $cov_{xy}$
of the two species, as a function of time for $\tau^c=100$. The
values of the multiplicative noise intensity are: $\sigma=10^{-12},
10^{-6}$, from top to bottom. The values of the other parameters and
the initial conditions are the same of Fig.~\ref{CML_series1}.}
 \label{CML_series3}
\end{figure}
Here $x^{(n)}_{i,j}$ and $y^{(n)}_{i,j}$ denote respectively the
densities of preys $x$ and $y$ in the site $(i,j)$ at the time step
$n$, $\mu$ is the growth rate and $D$ is the diffusion constant. The
interaction parameter $\beta^{(n)}$ corresponds to the value of
$\beta(t)$ taken at the time step $n$, according to
Eq.~(\ref{jump_rate}). $\zeta_{i,j}^{x(n)}$ and $\zeta_{i,j}^{y(n)}$
are independent Gaussian colored noise sources above defined in
Eq.~(\ref{colored_noise}). Finally $\sum_\gamma$ indicates the sum
over the four nearest neighbors. To evaluate the $1^{st}$ and
$2^{nd}$ order moments we define on the lattice, at the time step
$n$, the mean values $<x>_{_{CML}}^{(n)}$, $<y>_{_{CML}}^{(n)}$,
\begin{equation}
<z>_{_{CML}}^{(n)}~=~\frac{\sum_{i,j}z^{(n)}_{i,j}}{N}~, \qquad z =
x, y \label{CML moments}
\end{equation}
the standard deviations $var^{(n)}_x$, $var^{(n)}_y$
\begin{equation}
var^{(n)}_z =\sqrt{\frac{\sum_{i,j}
(z^{(n)}_{i,j}-<z>^{(n)})^2}{N}}~,  \qquad z = x, y \label{CML var}
\end{equation}
and the covariance $cov^{(n)}_{xy}$ of the two species
%
%
%
%
\begin{equation}
cov^{(n)}_{xy}=\frac{\sum_{i,j}(x^{(n)}_{i,j}-<x>^{(n)})
(y^{(n)}_{i,j}-<y>^{(n)})}{N},
 \label{covariance}
\end{equation}
where $N = 100\times 100$ is the number of lattice sites. We note
that $<x>_{_{CML}}$, $<y>_{_{CML}}$ and $var_x$, $var_y$,
$cov_{xy}$, corresponding respectively to $<x>$, $<y>$, and
$\mu_{20}$, $\mu_{02}$, $\mu_{11}$ obtained within the mean field
approach, are the $1^{st}$ and $2^{nd}$ order moments defined within
the scheme of the CML model. The time behavior of these quantities,
for three different values of the multiplicative noise intensity,
and for three different values of the correlation time is reported
in Figs.~\ref{CML_series1}-\ref{CML_series3}. Comparing these
results with those shown in
Figs.~\ref{moment_series1}-\ref{moment_series3}, we note that the
time behaviour obtained for the $1^{st}$ and $2^{nd}$ order moments
by the CML model is in a good qualitative agreement with those found
using the mean field approach. In particular we note the role played
by the multiplicative noise: higher noise intensity causes the
$2^{nd}$ order moments to increase (see in
Figs.~\ref{CML_series1}-\ref{CML_series3} the enhancement of the
oscillation maxima as the multiplicative noise intensity increases).
This indicates both a spread of the two species concentrations and a
spatial anticorrelation between them. These results agree with those
found within the formalism of the moment equations. We observe that
the time behaviour of $var_x$, $var_y$, $cov_{xy}$ is characterized
by the presence of oscillations whose maximum amplitude is reached
after a time delay. This peculiarity is the same of that found
within the mean field approach (see paragraph $3$). The
discrepancies in the oscillation intensities are due to: (i) the
Gaussian approximation in the moment formalism; (ii) the fact that
the species interaction in the CML model is restricted to the
nearest neighbors; (iii) the different stationary values of the
species densities in the considered models. Specifically, using for
$\beta$ an average value,
$\beta_{aver}=(\beta_{up}+\beta_{down})/2$, we get: $x_{st} = y_{st}
= 1/(1 + \beta_{aver}) \simeq 0.5$ for the mean field model, and
$x_{st}^{^{CML}} = y_{st}^{^{CML}} = (1 - 1/\mu)/(1 + \beta_{aver})
\simeq 0.25$ for the CML model.

\section{Conclusions}
\vskip-0.2cm

By using the moment formalism in Gaussian approximation, we describe
the time behavior of two competing species inside a two-dimensional
spatial domain in the presence both of a multiplicative colored
noise and a dichotomous noise. We find that the $1^{st}$ order
moments of the two species densities show correlated oscillations,
whose amplitude is independent on the multiplicative noise. However,
the behavior of the $2^{nd}$ order central moments depend strongly
both on the intensity $\sigma$ and the correlation time $\tau^c$ of
the multiplicative noise. In particular, the behavior of the
$2^{nd}$ order mixed moment $\mu_{11}$ indicates that higher values
of the multiplicative noise intensity push the two species towards
an anticorrelated regime characterized by oscillations whose maximum
amplitude is reached after a delay time: this delayed behaviour
depends on the correlation time $\tau^c$. We find a good qualitative
agreement between these results, obtained within the mean field
approach, and those found by the CML model. In view of some
applications of our model, to describe and to predict the behaviour
of biological species, we note that in real ecosystems the
fluctuations are characterized by a cut-off. Therefore experimental
data~\cite{Garcia}, whose dynamics is strongly affected by noisy
perturbations and stochastic environmental variables, can be better
modeled using sources of colored noise.
\section{Acknowledgments} \vskip-0.2cm
This work was supported by ESF (European Science Foundation)
STOCHDYN network, MIUR, INFM-CNR and CNISM.

\end{document}